# Opportunities for Machine Learning to Accelerate Halide Perovskite Commercialization and Scale-Up

Rishi E. Kumar[1,2], Armi Tiihonen[3], Shijing Sun[3], David P. Fenning[1,2]\*, Zhe Liu[3,4]\*, Tonio Buonassisi[3]\*

[1]Department of Nanoengineering, UC San Diego, La Jolla, CA 92093, USA

[2]Materials Science & Engineering Program, UC San Diego, La Jolla, CA 92093, USA

[3]Massachusetts Institute of Technology, Cambridge MA 02139, USA

[4]School of Materials Science & Engineering, Northwestern Polytechnical University, Xi'an, Shaanxi 710072, China

\*Correspondence to: dfenning@eng.ucsd.edu, zhe.liu@nwpu.edu.cn, buonassi@mit.edu

**Abstract**

While halide perovskites attract significant academic attention, examples of at-scale industrial production are still sparse. In this perspective, we review practical challenges hindering the commercialization of halide perovskites, and discuss how machine-learning (ML) tools could help: (1) active-learning algorithms that blend institutional knowledge and human expertise could help stabilize and rapidly update baseline manufacturing processes; (2) ML-powered metrology, including computer imaging, could help narrow the performance gap between large- and small-area devices; and (3) inference methods could help accelerate root-cause analysis by reconciling multiple data streams and simulations, focusing research effort on areas with highest probability for improvement. We conclude that to satisfy many of these challenges, incremental — not radical — adaptations of existing ML and statistical methods are needed. We identify resources to help develop in-house data-science talent, and propose how industry-academic partnerships could help adapt "ready-now" ML tools to specific industry needs, further improve process control by revealing underlying mechanisms, and develop "gamechanger" discovery-oriented algorithms to better navigate vast materials combination spaces and the literature.

**Introduction**

In under two decades, researchers propelled the AM1.5 efficiency of single-junction halide perovskite photovoltaic (PV) devices beyond 25% [1-4], in addition to demonstrating competitive performance in other optoelectronic applications including light-emitting diodes and photodetectors [5, 6]. However, few successful examples exist of commercial production at scale [7, 8]. Part of this is due to well-documented concerns surrounding perovskite instability [9, 10]. But conversations with industrial partners suggest that lesser-discussed and perhaps more mundane concerns exist, including but not limited to establishing a stable baseline process. This motivated the authors to engage in a broader and open-ended conversation with industry worldwide about the outstanding challenges they face.

Meanwhile, over the past decade, applied machine-learning (ML) methods have started to gain increased industrial acceptance, aiding complex process optimization and diagnosis [11-15]. During informal discussions with PV industry it became apparent that ML methods are slow to gain acceptance, in part because of concerns surrounding interpretability, talent accessibility, and uncertain advantages over traditional design of experiments.

These observations motivate the central questions of this Perspective: (1) What are the key barriers that industry faces to commercialize halide-perovskite optoelectronics, and (2) Can ML methods assist industry to overcome these challenges? We hope this Perspective motivates future work, especially in the field of applied ML, aiding perovskites to "see the light of day."

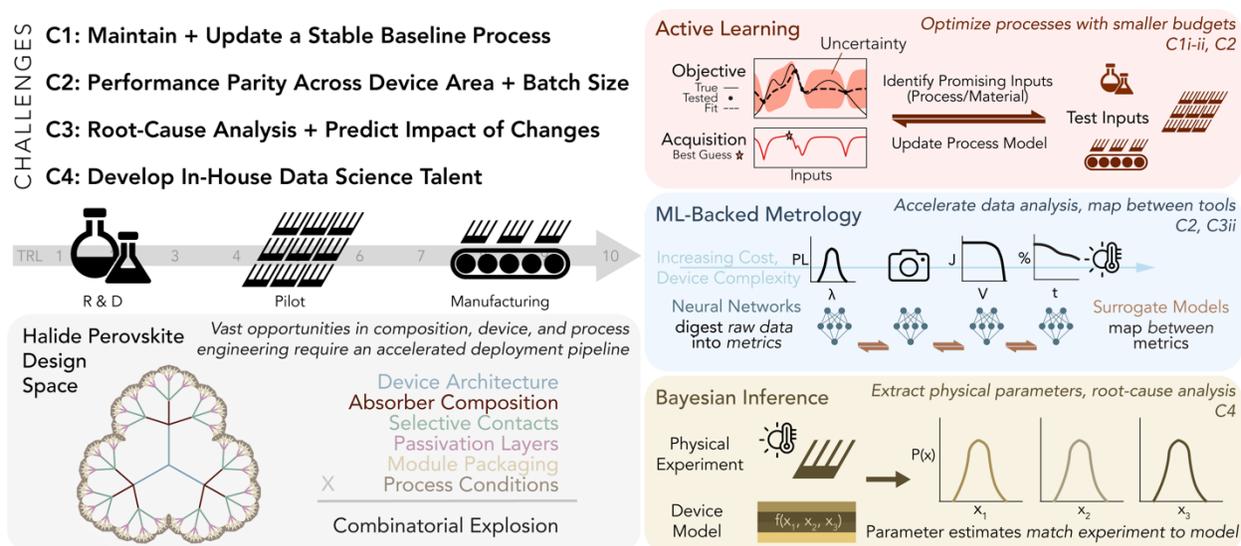

**Figure 1**. Schematic of how "ready-now" machine learning tools can assist perovskite device commercialization. Techniques are labeled by the sections in which they are discussed within the text.

**What Challenges does Industry Face to Manufacture Perovskites at Scale?**

Through discussions with industry worldwide, we observe a continuum of concerns spanning efforts across the Technology Readiness Levels (TRLs) [16]. These concerns reach from the "process-development phase" (developing a new process in R&D and translating it to small lots in production, roughly TRLs 5–7) to the later "manufacturing scale-up phase" (achieving high performance in qualification runs, roughly TRLs 7–9). We distill and articulate industry needs in the following challenges:

C1. Maintaining a stable baseline process (as measured by efficiency and environmental stability), and frequently updating the process to incorporate the latest advances in the public domain
C2. Achieving large-area devices and/or large-scale manufacturing that perform close to small area devices
C3. Troubleshooting root causes of under-performance (ideally in-line and early in the manufacturing process), and estimating performance improvements arising from process or device-architecture modifications
C4. Developing in-house data-science talent

While most academic research to date on halide perovskites has focused on earlier TRLs, the translational activities that are required for perovskites to reach commercial readiness include substantial scientific unknowns regarding perovskite processing, where academic-industrial collaborations in new directions to directly address remaining challenges may prove fruitful. Furthermore, the academic research community has applied ML to other manufacturing systems [17-20] that directly address some of the needs above; successful examples of industrial ML scale-up include battery lifetime prediction for electric vehicles [21, 22], medicinal diagnostics [23], and high-strength structural materials [24, 25]. The next section explores opportunities to apply these learnings to perovskite commercialization.

**How Can ML Methods Accelerate Perovskite Industrialization?**

Here we describe ML tools that may aid perovskite manufacturing and commercialization, using the aforementioned challenges as section headers. As industry places a premium in risk reduction, we emphasize *existing* ML-based approaches, with "existence proofs" in either perovskites or manufacturing in general. In the last section on "gamechangers," we describe how ongoing academic work applying ML to perovskites connects to industry needs. For an introduction to ML methods applied to materials research, the reader is referred to recent reviews [3, 26-31].

1. *Identifying & maintaining stable baselines that can be rapidly upgraded*

It can be challenging to identify and maintain a stable perovskite baseline. Perovskites are sensitive to multiple spatially and temporally varying process parameters, including environmental conditions. The multinary nature of the bulk composition, as well as sensitivity to commonly uncontrolled variables (*e.g.*, residual atmospheric solvent vapor pressure and relative humidity [32], among others), can affect product quality. We consider how *established* ML methods may be able to assist in identifying and maintaining more stable baselines, especially when combined with stringent process control to minimize variance. Open-source development is becoming common practice in the field, where code is often made available with licenses that offer users the right to build foreground IP.

*i. Find the baseline*. Active learning approaches, including Bayesian optimization (BO, [33-

36]), recently gained popularity to identify optimal perovskite processing conditions with small experimental budgets. A recent pre-print by Liu *et al.* applied BO to optimize perovskite solar cell devices synthesized at Stanford by rapid spray-plasma processing (RSPP) [37]. They demonstrate advantages over traditional design of experiments (DOE) approaches (including factorial sampling with partial grid search and one-variable-at-a-time search), identifying multiple process conditions with higher performance after five batches of optimization. Sun *et al.* [38] used a related BO-based approach to identify a range of spin-coated perovskite absorber compositions with optimum environmental stability, using only 1.8% of available compositions. Active learning has also been demonstrated to be effective in guiding high-throughput computational screening of candidate perovskite materials [39, 40].

Active-learning methods have been applied to other device elements and related absorbers to identify baseline processing conditions that maximize conversion efficiency and reaction yields. BO has been used to guide experimental search for fabrication conditions to maximize the conductivity of hole-transport materials [41]. Langner *et al.* [42] showed that BO of an organic photovoltaic (OPV) absorber layer defined in a quaternary compositional space takes 30x fewer samples than brute-force combinatorics. As a point of comparison, a similar reduction in the number of runs can be difficult to achieve in a high-resolution fractional factorial design (*e.g.* 1/32 fractionation) without significant *a priori* knowledge and confidence about which factors are suitable to confound. Several studies combined ML and robotics to autonomously fabricate and optimize perovskite single crystals, thin film photoabsorbers, quantum dots, and the charge transport materials for perovskite solar devices [41, 43-46]. BO has also been shown to be an effective search algorithm for optimization of chemical reaction yields, where the state-of-the-art yields for reactions defined in search spaces of $10^5$ points were surpassed with experimental budgets of $10^1$–$10^2$ points [47]. The power of active learning methods to reduce the number of experiments required to advance a figure of merit is increasingly well demonstrated and quite general.

Recent extensions to BO include multi-objective optimization to map Pareto fronts [48-50], benchmarking work to optimize the degree of exploration and exploitation of the BO algorithm for experimental materials science applications [51-54], dimensionality-reduction approaches to address design spaces >20 input variables [42, 47, 55-58], and uncertainty quantification [59-61]. In summary, the literature has consistently shown that active learning-driven optimization typically requires a total experimental budget of 10 to 100 unique sampling conditions – well within the usual throughput of an average team, and often superior to human-driven experimental design [41, 42, 47, 62-64].

*ii. Maintain the baseline*. Recent BO studies applied to manufacturing focus on identifying not only the highest performing, but the most robust process. This can be achieved by identifying relatively flat plateaus in the performance parameter of merit, *e.g.*, using the open-source Golem algorithm [65]. To identify process drifts on the fly and predict ultimate performance, ML applied to luminescence imaging on semi-fabricates was used for silicon solar cells [66, 67]. Periodic ML-driven sampling may nevertheless be needed to adjust baselines to slowly fluctuating uncontrolled variables, such as seasonal variations in ambient humidity that was demonstrated to affect perovskite crystal growth [32].

*iii. Upgrade the baseline*. The target compositions and manufacturing processes of various device layers evolve rapidly, thanks to both internal and external R&D, but are typically challenging to incorporate into a baseline process. The ability to upgrade a baseline process is therefore of interest. Liu *et al*. report using prior machine settings as a "soft constraint" within a Bayesian optimization framework, allowing the search for the new optimum to leverage old information as a "warm start" [37]. Sun *et al*. [38] and Gongora *et al*. [68] use first-principles simulations in a similar way.

2. *Scale-up: Achieving large-area devices and/or large-scale manufacturing that perform close to small area*

In the early days of perovskite development, increasing device area resulted in efficiency declines over twice as large as those in commercial inorganic PV materials [7, 69]. Recent lab-scale demonstrations have reduced these losses [70], although achieving high efficiencies with large active areas using manufacturing-friendly processes [71] remains a challenge.

Computer vision offers rapid identification of spatial and temporal inhomogeneities. When computer vision is combined with ML, the performance impacts of the aforementioned inhomogeneities can be predicted, and process control corrections can be suggested [72]. Such metrology can be applied across TRLs, accumulating training data and building efficacy as they are moved from benchtop investigation to in-line process monitoring. Well-established imaging-processing tools leverage the benefits of ML in handling large datasets of pixels and therefore provide an effective avenue in monitoring, quantifying, and controlling spatially varying parameters in large-area devices. Photoluminescence and electroluminescence images, coupled to device simulations and machine learning, enable in-line performance prediction for inorganic PV [73-76] and have recently been extended to perovskites [61, 77]. ML algorithms developed by Tian *et al.* [78] and Taherimakhsousi *et al*. [79] estimate film thickness and defect density non-destructively by processing optical information. These ML-enabled methods augment (rather than substitute) existing efforts to tighten process controls, identify anomalies, identify uncontrolled variables, and reduce spatial and temporal non-uniformities, potentially enabling improved performance prediction earlier in the manufacturing line, early detection of process excursions and quality incidents, and more effective preventative maintenance.

3. *Troubleshooting root causes of under-performance (ideally in-line and early in the manufacturing process), and estimating performance improvements arising from process or device-architecture modifications*

*i. Extract physical parameters of solar cells*. Root-cause(s) of underperformance of solar cells can be extracted by pairing physics-based simulations with current-voltage curve measurements. However, this method of troubleshooting can be time-consuming and inconclusive, as the number of simulation fitting variables can be vast, and two or more limiting factors often combine to limit performance.

Bayesian inference is a probabilistic method to combine models and measurements in a statistically rigorous way, representing as probability distributions the underlying variables and their correlations. Brandt *et al.*[80] and Kurchin *et al*. [81] demonstrated the application of Bayesian inference to underlying parameter extraction (e.g., minority-carrier lifetime, surface recombination velocity, interfacial energy barriers, bulk-defect properties…) for thin film and silicon solar cells. Ren, Oviedo *et al.* [82] coupled Bayesian inference to a heuristic process model to design a time-temperature profile for a gallium arsenide (GaAs) solar cell device stack, improving device efficiency by 6.5% relative to a DOE best; essentially, the ML algorithm both optimized and identified root cause(s) of underperformance, informing the user how to improve the process. Oviedo applied Bayesian inference to extract the root-cause(s) of degradation during environmental testing of organic PV devices [83]. One of the advantages of this method is to decouple efficiency contributors in a finished device stack, where all device layers and interfaces are in their final state. A key enabler of these inference methods, is that a neural network "surrogate" model can be trained to mimic a numerical device simulator over a range of input values, running 100–1000x faster than the original simulator (see supporting information Fig. S1 in [82]). This approach enables even complex device simulators to be incorporated into Bayesian inference algorithms, inasmuch as the surrogate model's training data spans the entire range of each inferred output with sufficient resolution.

The challenges for wider adoption of Bayesian inference to emerging materials like perovskites are: (i) To produce perovskite device-physics models that capture all of the performance-relevant underlying physics, including polarization, ion migration, and second-phase formation [84-87]. (ii) To adapt Bayesian inference models to high-efficiency devices with incremental improvements. To date, device efficiencies <20% have been studied, with large efficiency spreads between samples. In manufacturing, higher efficiencies and smaller spreads are more typical; the smaller signal-to-noise means both more accurate model and higher-quality data are required, *e.g.*, combining *J-V* measurements with other device measurements inside a Bayesian inference framework.

*ii. Find early predictors for ultimate performance*. ML models can help engineers establish correlations between characterization results within the fabrication or stability testing process as well as ultimate performance parameters of merit, saving time and resources. As an R&D example, optical images on bare perovskite films (which turn yellow during degradation) can give an indication of the environmental stability of finished perovskite devices [88, 89]; this information can be combined with an active learning algorithm to guide compositional tuning in perovskites, increasing device stability by ~2x within the ternary [Cs-MA-FA]-Pb-I compositional space with only a small fraction of this space fabricated into full devices [38]. For organic solar cells, Oviedo [83] developed a time-series forecast model that estimates device efficiency using only the first few hours (*e.g.*, the initial 5–10% of total testing time), significantly shortening the time required for degradation testing. For silicon solar cells, Kunze *et al.* [67] established an ML model to predict the final *I-V* metrics (*i.e.*, $J_{SC}$, $V_{OC}$, *FF*, efficiency) based on luminescence images collected on the production line. Stoddard *et al.* [90] and Howard *et al.* [91] applied recurrent neural networks to predict the photoluminescence dynamics of lead-halide perovskite thin films under humidity cycling, demonstrating the value of ML in performing effective time series forecasting for perovskite optoelectronic behaviors. Despite these encouraging examples, we are currently at an early stage in predicting the operational reliability of perovskite solar cells, demanding close collaborations between device fabrication and long-term performance testing, and seamless integration between physical and data-driven models.

## 4. *Developing in-house data-science talent*

In the late 1990's, when website design required coding, a high premium was paid to hire rare HTML talent. As graphical user interfaces (GUIs) evolved throughout the 2000's, basic website-making became accessible to anyone with a computer, while professionals specialized in higher-end services. ML is currently at a similar early stage in its technology adoption cycle, with many companies and open-source communities launching software-as-a-service products. As ML code libraries, and GUI wrappers around ML libraries, become more commonplace, and a dominant design emerges, it is foreseeable that basic ML capabilities will become as accessible as word processors are today.

Navigating this early phase of ML adoption can be challenging for perovskite manufacturers, and it can be tempting just to "wait it out" [92]. But not upskilling existing staff in ML methods carries not only opportunity cost, but also timeline risk, because unfamiliarity breeds mistrust that delays ML adoption. Here are some resources for the busy employee and manager to upskill in applied ML:

- Applied ML tutorials and repositories of practical examples are increasingly available. Examples include MLOps [93] and Accelerated Materials Development for Manufacturing [94].
- The above examples often include open-source code on repositories like GitHub or Papers with Code. These codes and libraries can often be adapted to your problem. Stack Overflow is a most useful community for troubleshooting.
- To keep abreast of literature, the authors prefer digests that access arXiv pre-prints (which are becoming common practice in applied ML), such as Google Scholar. Social media is also a surprisingly effective distribution channel for pre-prints; Twitter has an active computational chemistry (#compchem) community, and a LinkedIn group by Benji Maruyama on Autonomous Research Systems [95] filters many relevant papers. Peer review has new embodiments when code and datasets can be locally tested within minutes.

The tools described here represent the authors' perspective as of 2021-Q4. This is a fast-evolving field. We are likely in the "fluid" phase described by Utterback [96], where a variety of ML technology options exist prior to convergence on a dominant design. In this phase, adaptiveness is essential. For example, one of the authors recently upgraded their group's BO capabilities from GPyOpt (development and maintenance ended in 2020 [97]) to BoTorch [98], which required a quarter-long full-time employee investment.

## 5. *Gamechangers*

Ongoing academic research into applied ML, if successful, could meaningfully aid perovskite commercialization, providing new and potentially transformative tools.

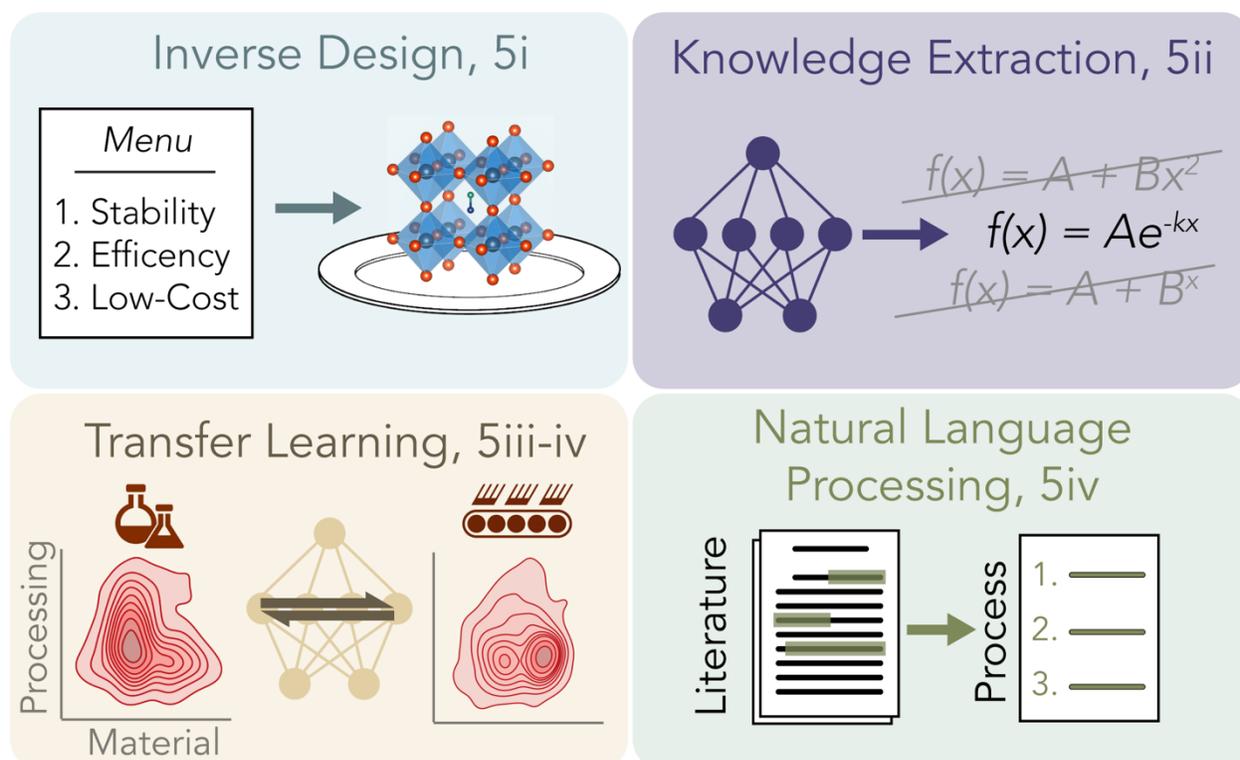

**Figure 2**: "Game changing" ML techniques for photovoltaic research, development, and deployment. The sections in which these methods are described are labeled in each bubble.

*i. Inverse design tools capable of solving real-world problems.* Successful inverse design capabilities could enable the discovery of new perovskite composition spaces that achieve improved environmental stability, sustained high performance over a wider range of illumination conditions (*e.g.*, for tandems or non-AM1.5 operating conditions), removal of lead, improved resilience to temporal and spatial variance in manufacturing, or some combination thereof. Most experiments and simulation tools use materials and process variables as inputs and calculate "properties" as outputs. Solving the "inverse problem" involves designing an algorithm that instead uses *desired properties as inputs* to generate a prospective material and process — even suggesting materials not included in existing perovskite databases [99, 100] or DFT calculations [101]. While this vision was formulated for some time [102, 103] the rise of variational autoencoders (VAE) enabled researchers to encode property-process-structure relationships within a generative model [104-107]. Related approaches include using generative adversarial networks (GAN) or genetic algorithms (GA) to generate candidate compounds through directed evolution. Outstanding challenges include filtering out candidate materials that cannot be synthesized [108-112], recommending synthesis pathways for those that can [106, 113], and coupling inverse design algorithms to materials-optimization platforms to experimentally test new candidates.

*ii. Extracting physical insights from ML models.* Successful "knowledge extraction" – derivation of greater physical insights- from ML models could enable both greater acceptance of ML methods by human researchers and greater generalization of results. This topic constitutes the ML subfields of Explainable Artificial Intelligence and Scientific ML [114-116]. A straightforward method uses ML to "fit" unknown input parameters of simulations [117-119], with similar advantages and disadvantages as Bayesian inference. Knowledge can also be extracted from trained ML algorithms in the form of feature importance ranking, *e.g.*, Shapley Additive Explanations (SHAP, [120, 121]) and Local Interpretable Model-agnostic Explanations (LIME, [122]), which have been used to illuminate how input materials and process parameters affect perovskite environmental stability [45, 88]; however, they must be applied with caution to avoid user bias [123, 124]. ML methods that construct decision trees [125] are inherently interpretable, exemplified by Kong *et al.*'s [126] reduction of materials databases into design rules to determine inorganic crystal structures. Scientific ML is an emergent class of models with the objective of finding an equation (often from a library of plausible functional forms) that best describes the relationship between data inputs and outputs. Naik *et al.* [127] used scientific ML to determine a rate equation for degradation of $MAPbI_3$ films, the form of which suggested that the film degradation proceeds by an autocatalytic reaction.

*iii. Facilitating process transfer between sites, equipment, and operators*. The replication of perovskite syntheses across even operators within the same laboratory has proven notoriously difficult. If a synthesis or manufacturing process can be captured via an ML model, transfer learning may provide a means to rapidly tune the process to a new setting. If successful, transfer learning could accelerate recipe adaptation between manufacturing lines or within the same manufacturing line over time (longitudinally), transfer of new processes from R&D into manufacturing, and possibly between manufacturing methods (*e.g.*, spin coating to chemical vapour deposition).

*iv. Automatically digesting the literature into the research and development pipeline*. New materials and processes are constantly being described in the perovskite literature. Advances in natural language processing (NLP) could facilitate programmatic extraction and deployment of freshly published processes into the R&D pipeline, augmenting all previously-mentioned tools presented in this article. While such a platform remains hypothetical for perovskite research, Mehr *et al.* [128] have demonstrated a nearly autonomous workflow for robotic execution of chemical syntheses mined from the literature. The NLP toolset required to distill the literature is ever-growing, with text-mining tools like ChemDataExtractor [129] and ChemicalTagger [130] and image-mining tools like EXSCLAIM! [131] (EXtraction, Separation, and Caption-based natural Language Annotation of IMages) currently available for download. New candidate materials or recipes could be explored using a combination of voting rules, transfer learning, and active learning to rapidly test high-probability candidates and automatically incorporate innovations into manufacturing lines.

**Conclusions**

The exquisite sensitivity of perovskites to local composition and process variables, the vastness of possibilities for single layers and layer combinations, the rapid expansion of the academic literature, and the pressing need to deliver technologies for sustainability encourage the adoption of an R&D framework that arbitrages the significant advances made in ML. ML algorithms hold promise to rapidly and adaptively downselect and optimize complex design spaces for technoeconomically relevant outputs. We present a summary of "ready-now" ML algorithms that may aid industrial perovskite device development. We posit that in many cases, ML adoption is now an incremental advancement, within reach of many industry teams, with code and datasets freely available online. To help "bridge the gap," we advocate for greater interaction between academic and industry teams concerning the topic of Applied ML.

**Conflict of Interest Declaration**

Two of the authors (Z.L. and T.B.) own equity in a startup, Xinterra, focused on applying ML to develop novel materials.

**Acknowledgements**

The authors thank their industrial contacts, group members, and collaborators for numerous helpful discussions that shaped this perspective.

**References:**

[1] D. Li, D. Zhang, K.-S. Lim, Y. Hu, Y. Rong, A. Mei, N.-G. Park, and H. Han, "A Review on Scaling Up Perovskite Solar Cells," *Advanced Functional Materials,* vol. 31, no. 12, p. 2008621, 2021, doi: https://doi.org/10.1002/adfm.202008621.

[2] C. A. R. Perini, T. A. S. Doherty, S. D. Stranks, J.-P. Correa-Baena, and R. L. Z. Hoye, "Pressing challenges in halide perovskite photovoltaics—from the atomic to module level," *Joule,* vol. 5, no. 5, pp. 1024-1030, 2021, doi: https://doi.org/10.1016/j.joule.2021.03.011.

[3] Q. Tao, P. Xu, M. Li, and W. Lu, "Machine learning for perovskite materials design and discovery," *npj Computational Materials,* vol. 7, no. 1, 2021, doi: 10.1038/s41524-021-00495-8.

[4] "NREL Champion Photovoltaic Module Efficiency Chart (rev. 05/10/2021)." https://www.nrel.gov/pv/assets/pdfs/champion-module-efficiencies.pdf, accessed 10/06/2021,

[5] S. A. Veldhuis, P. P. Boix, N. Yantara, M. Li, T. C. Sum, N. Mathews, and S. G. Mhaisalkar, "Perovskite Materials for Light-Emitting Diodes and Lasers," *Advanced Materials,* vol. 28, no. 32, pp. 6804-6834, 2016, doi: 10.1002/adma.201600669.

[6] H. Wu, Y. Ge, G. Niu, and J. Tang, "Metal halide perovskites for X-ray detection and imaging," *Matter,* vol. 4, no. 1, pp. 144-163, 2021, doi: 10.1016/j.matt.2020.11.015.

[7] A. Extance, "The reality behind solar power's next star material," *Nature,* vol. 570, no. 7762, pp. 429-432, 2019, doi: 10.1038/D41586-019-01985-Y.

[8] C. Crownhart. "Can the most exciting new solar material live up to its hype? ." MIT Technology Review. https://www.technologyreview.com/2021/06/29/1027451/perovskite-solar-panels-hype-commercial-debut/, accessed 10/06/2021,

[9] M. V. Khenkin, E. A. Katz, A. Abate, G. Bardizza, J. J. Berry, C. Brabec, F. Brunetti, V. Bulović, Q. Burlingame, A. Di Carlo, R. Cheacharoen, Y.-B. Cheng, A. Colsmann, S. Cros, K. Domanski, M. Dusza, C. J. Fell, S. R. Forrest, Y. Galagan, D. Di Girolamo, M. Grätzel, A. Hagfeldt, E. von Hauff, H. Hoppe, J. Kettle, H. Köbler, M. S. Leite, S. Liu, Y.-L. Loo, J. M. Luther, C.-Q. Ma, M. Madsen, M. Manceau, M. Matheron, M. McGehee, R. Meitzner, M. K. Nazeeruddin, A. F. Nogueira, Ç. Odabaşı, A. Osherov, N.-G. Park, M. O. Reese, F. De Rossi, M. Saliba, U. S. Schubert, H. J. Snaith, S. D. Stranks, W. Tress, P. A. Troshin, V. Turkovic, S. Veenstra, I. Visoly-Fisher, A. Walsh, T. Watson, H. Xie, R. Yıldırım, S. M. Zakeeruddin, K. Zhu, and M. Lira-Cantu, "Consensus statement for stability assessment and reporting for perovskite photovoltaics based on ISOS procedures," *Nature Energy,* vol. 5, no. 1, pp. 35-49, 2020, doi: 10.1038/s41560-019-0529-5.

[10] S. P. Dunfield, L. Bliss, F. Zhang, J. M. Luther, K. Zhu, M. F. A. M. van Hest, M. O. Reese, and J. J. Berry, "From Defects to Degradation: A Mechanistic Understanding of Degradation in Perovskite Solar Cell Devices and Modules," *Advanced Energy Materials,* vol. 10, no. 26, p. 1904054, 2020, doi: 10.1002/aenm.201904054.

[11] T. P. Carvalho, F. A. A. M. N. Soares, R. Vita, R. d. P. Francisco, J. P. Basto, and S. G. S. Alcalá, "A systematic literature review of machine learning methods applied to predictive maintenance," *Computers & Industrial Engineering,* vol. 137, p. 106024, 2019, doi: 10.1016/j.cie.2019.106024.

[12] D. Weichert, P. Link, A. Stoll, S. Rüping, S. Ihlenfeldt, and S. Wrobel, "A review of

[12] machine learning for the optimization of production processes," *The International Journal of Advanced Manufacturing Technology,* vol. 104, no. 5-8, pp. 1889-1902, 2019, doi: 10.1007/s00170-019-03988-5.

[13] N. Savage, "Tapping into the drug discovery potential of AI," *Biopharma Dealmakers,* 2021

[14] M. D. Abràmoff, P. T. Lavin, M. Birch, N. Shah, and J. C. Folk, "Pivotal trial of an autonomous AI-based diagnostic system for detection of diabetic retinopathy in primary care offices," *npj Digital Medicine,* vol. 1, no. 1, p. 39, 2018, doi: 10.1038/s41746-018-0040-6.

[15] H. Patel, D. Prajapati, D. Mahida, and M. Shah, "Transforming petroleum downstream sector through big data: a holistic review," *Journal of Petroleum Exploration and Production Technology,* vol. 10, no. 6, pp. 2601-2611, 2020, doi: 10.1007/s13202-020-00889-2.

[16] J. Mankins, "Technology Readiness Level – A White Paper," Office of Space Access and Technology, NASA, 1995. [Online]. Available: www.researchgate.net/publication/247705707_Technology_Readiness_Level_-_A_White_Paper

[17] S. Mäkinen, H. Skogström, E. Laaksonen, and T. Mikkonen, "Who Needs MLOps: What Data Scientists Seek to Accomplish and How Can MLOps Help?," 2021, arXiv:2103.08942. [Online] Available: arxiv.org/abs/2103.08942

[18] L. Miranda. "Towards data-centric machine learning: a short review." jvmiranda921.github.io/notebook/2021/07/30/data-centric-ml/, accessed 10/6/2021,

[19] T. Wuest, C. Irgens, and K.-D. Thoben, "An approach to quality monitoring in manufacturing using supervised machine learning on product state data," *Journal of Intelligent Manufacturing,* vol. 25, pp. 1167-1180, 2014, doi: 10.1007/s10845-013-0761-y.

[20] L. Monostori, A. Márkus, and H. Brussel, "Machine Learning Approaches to Manufacturing," *CIRP Annals - Manufacturing Technology,* vol. 45, 1996, doi: 10.1016/S0007-8506(18)30216-6.

[21] M. Aykol, P. Herring, and A. Anapolsky, "Machine learning for continuous innovation in battery technologies," *Nature Reviews Materials,* vol. 5, no. 10, pp. 725-727, 2020, doi: 10.1038/s41578-020-0216-y.

[22] K. A. Severson, P. M. Attia, N. Jin, N. Perkins, B. Jiang, Z. Yang, M. H. Chen, M. Aykol, P. K. Herring, D. Fraggedakis, M. Z. Bazant, S. J. Harris, W. C. Chueh, and R. D. Braatz, "Data-driven prediction of battery cycle life before capacity degradation," *Nature Energy,* vol. 4, no. 5, pp. 383-391, 2019, doi: 10.1038/s41560-019-0356-8.

[23] E. J. Topol, "High-performance medicine: the convergence of human and artificial intelligence," *Nature Medicine,* vol. 25, no. 1, pp. 44-56, 2019, doi: 10.1038/s41591-018-0300-7.

[24] H. Zhuang, "From evidence to new high-entropy alloys," *Nature Computational Science,* vol. 1, no. 7, pp. 458-459, 2021, doi: 10.1038/s43588-021-00100-4.

[25] J. H. Martin, B. D. Yahata, J. M. Hundley, J. A. Mayer, T. A. Schaedler, and T. M. Pollock, "3D printing of high-strength aluminium alloys," *Nature,* vol. 549, no. 7672, pp. 365-369, 2017, doi: 10.1038/nature23894.

[26] J. Li, K. Lim, H. Yang, Z. Ren, S. Raghavan, P.-Y. Chen, T. Buonassisi, and X. Wang, "AI Applications through the Whole Life Cycle of Material Discovery," *Matter,* vol. 3,

no. 2, pp. 393-432, 2020, doi: 10.1016/j.matt.2020.06.011.

[27] K. T. Butler, D. W. Davies, H. Cartwright, O. Isayev, and A. Walsh, "Machine learning for molecular and materials science," *Nature,* vol. 559, no. 7715, pp. 547-555, 2018, doi: 10.1038/s41586-018-0337-2.

[28] J.-P. Correa-Baena, K. Hippalgaonkar, J. van Duren, S. Jaffer, V. R. Chandrasekhar, V. Stevanovic, C. Wadia, S. Guha, and T. Buonassisi, "Accelerating Materials Development via Automation, Machine Learning, and High-Performance Computing," *Joule,* vol. 2, no. 8, pp. 1410-1420, 2018, doi: 10.1016/j.joule.2018.05.009.

[29] D. P. Tabor, L. M. Roch, S. K. Saikin, C. Kreisbeck, D. Sheberla, J. H. Montoya, S. Dwaraknath, M. Aykol, C. Ortiz, H. Tribukait, C. Amador-Bedolla, C. J. Brabec, B. Maruyama, K. A. Persson, and A. Aspuru-Guzik, "Accelerating the discovery of materials for clean energy in the era of smart automation," *Nature Reviews Materials,* vol. 3, no. 5, pp. 5-20, 2018, doi: 10.1038/s41578-018-0005-z.

[30] E. Stach, B. DeCost, A. G. Kusne, J. Hattrick-Simpers, K. A. Brown, K. G. Reyes, J. Schrier, S. Billinge, T. Buonassisi, I. Foster, C. P. Gomes, J. M. Gregoire, A. Mehta, J. Montoya, E. Olivetti, C. Park, E. Rotenberg, S. K. Saikin, S. Smullin, V. Stanev, and B. Maruyama, "Autonomous experimentation systems for materials development: A community perspective," *Matter,* vol. 4, no. 9, pp. 2702-2726, 2021, doi: 10.1016/j.matt.2021.06.036.

[31] S. Kalinin and et al., *in press,* 2021.

[32] P. W. Nega, Z. Li, V. Ghosh, J. Thapa, S. Sun, N. T. P. Hartono, M. A. N. Nellikkal, A. J. Norquist, T. Buonassisi, E. M. Chan, and J. Schrier, "Using automated serendipity to discover how trace water promotes and inhibits lead halide perovskite crystal formation," *Applied Physics Letters,* vol. 119, no. 4, p. 041903, 2021, doi: 10.1063/5.0059767.

[33] V. L. Deringer, A. P. Bartók, N. Bernstein, D. M. Wilkins, M. Ceriotti, and G. Csányi, "Gaussian Process Regression for Materials and Molecules," *Chemical Reviews,* vol. 121, no. 16, pp. 10073-10141, 2021, doi: 10.1021/acs.chemrev.1c00022.

[34] J. Chang, P. Nikolaev, J. Carpena-Núñez, R. Rao, K. Decker, A. E. Islam, J. Kim, M. A. Pitt, J. I. Myung, and B. Maruyama, "Efficient Closed-loop Maximization of Carbon Nanotube Growth Rate using Bayesian Optimization," *Scientific Reports,* vol. 10, no. 1, p. 9040, 2020, doi: 10.1038/s41598-020-64397-3.

[35] Y. Zhang, D. W. Apley, and W. Chen, "Bayesian Optimization for Materials Design with Mixed Quantitative and Qualitative Variables," *Scientific Reports,* vol. 10, no. 1, p. 4924, 2020, doi: 10.1038/s41598-020-60652-9.

[36] F. Häse, M. Aldeghi, R. J. Hickman, L. M. Roch, and A. Aspuru-Guzik, "Gryffin: An algorithm for Bayesian optimization of categorical variables informed by expert knowledge," *Applied Physics Reviews,* vol. 8, no. 3, 2021, doi: 10.1063/5.0048164.

[37] Z. Liu, N. Rolston, A. C. Flick, T. Colburn, Z. Ren, R. H. Dauskardt, and T. Buonassisi, "Machine learning with knowledge constraints for process optimization of open-air perovskite solar cell manufacturing," 2021, arXiv:2110.01387. [Online] Available: arxiv.org/abs/2110.01387

[38] S. Sun, A. Tiihonen, F. Oviedo, Z. Liu, J. Thapa, Y. Zhao, N. T. P. Hartono, A. Goyal, T. Heumueller, C. Batali, A. Encinas, J. J. Yoo, R. Li, Z. Ren, I. M. Peters, C. J. Brabec, M. G. Bawendi, V. Stevanovic, J. Fisher, and T. Buonassisi, "A data fusion approach to optimize compositional stability of halide perovskites," *Matter,* vol. 4, no. 4, pp. 1305-1322, 2021, doi: 10.1016/j.matt.2021.01.008.

[39] H. C. Herbol, W. Hu, P. Frazier, P. Clancy, and M. Poloczek, "Efficient search of compositional space for hybrid organic–inorganic perovskites via Bayesian optimization," *npj Computational Materials,* vol. 4, no. 1, 2018, doi: 10.1038/s41524-018-0106-7.

[40] X. Chen, C. Wang, Z. Li, Z. Hou, and W.-J. Yin, "Bayesian optimization based on a unified figure of merit for accelerated materials screening: A case study of halide perovskites," *Science China Materials,* vol. 63, no. 6, pp. 1024-1035, 2020, doi: 10.1007/s40843-019-1255-4.

[41] B. P. MacLeod, F. G. L. Parlane, and T. D. Morrissey, "Self-driving laboratory for accelerated discovery of thin-film materials," *Science Advances,* vol. 6, p. eaaz8867, 2020.

[42] S. Langner, F. Hase, J. D. Perea, T. Stubhan, J. Hauch, L. M. Roch, T. Heumueller, A. Aspuru-Guzik, and C. J. Brabec, "Beyond ternary OPV: high-throughput experimentation and self-driving laboratories optimize multicomponent systems," *Advanced Materials,* vol. 32, no. 14, p. e1907801, 2020, doi: 10.1002/adma.201907801.

[43] R. W. Epps, M. S. Bowen, A. A. Volk, K. Abdel-Latif, S. Han, K. G. Reyes, A. Amassian, and M. Abolhasani, "Artificial Chemist: An Autonomous Quantum Dot Synthesis Bot," *Advanced Materials,* vol. 32, no. 30, p. 2001626, 2020, doi: 10.1002/adma.202001626.

[44] Z. Li, M. A. Najeeb, L. Alves, A. Z. Sherman, V. Shekar, P. Cruz Parrilla, I. M. Pendleton, W. Wang, P. W. Nega, M. Zeller, J. Schrier, A. J. Norquist, and E. M. Chan, "Robot-Accelerated Perovskite Investigation and Discovery," *Chemistry of Materials,* vol. 32, no. 13, pp. 5650-5663, 2020, doi: 10.1021/acs.chemmater.0c01153.

[45] Y. Zhao, J. Zhang, Z. Xu, S. Sun, S. Langner, N. T. P. Hartono, T. Heumueller, Y. Hou, J. Elia, N. Li, G. J. Matt, X. Du, W. Meng, A. Osvet, K. Zhang, T. Stubhan, Y. Feng, J. Hauch, E. H. Sargent, T. Buonassisi, and C. J. Brabec, "Discovery of temperature-induced stability reversal in perovskites using high-throughput robotic learning," *Nature Communications,* vol. 12, no. 1, pp. 2191-2191, 2021, doi: 10.1038/s41467-021-22472-x.

[46] J. Kirman, A. Johnston, D. A. Kuntz, M. Askerka, Y. Gao, P. Todorović, D. Ma, G. G. Privé, and E. H. Sargent, "Machine-learning-accelerated perovskite crystallization," *Matter,* vol. 2, no. 4, pp. 938-947, 2020, doi: 10.1016/j.matt.2020.02.012.

[47] B. J. Shields, J. Stevens, J. Li, M. Parasram, F. Damani, J. I. M. Alvarado, J. M. Janey, R. P. Adams, and A. G. Doyle, "Bayesian reaction optimization as a tool for chemical synthesis," *Nature,* vol. 590, no. 7844, pp. 89-96, 2021, doi: 10.1038/s41586-021-03213-y.

[48] B. P. MacLeod, F. G. L. Parlane, K. E. Dettelbach, M. S. Elliott, C. C. Rupnow, T. D. Morrissey, T. H. Haley, O. Proskurin, M. B. Rooney, N. Taherimakhsousi, D. J. Dvorak, H. N. Chiu, C. E. B. Waizenegger, K. Ocean, and C. P. Berlinguette, "Advancing the Pareto front using a self-driving laboratory," 2021, arXiv:2106.08899 [Online] Available: arxiv.org/abs/2106.08899

[49] T. Erps, M. Foshey, M. K. Luković, W. Shou, H. H. Goetzke, H. Dietsch, K. Stoll, B. v. Vacano, and W. Matusik, "Accelerated Discovery of 3D Printing Materials Using Data-Driven Multi-Objective Optimization," 2021, arXiv:2106.15697. [Online] Available: arxiv.org/abs/2106.15697

[50] A. M. Schweidtmann, A. D. Clayton, N. Holmes, E. Bradford, R. A. Bourne, and A. A. Lapkin, "Machine learning meets continuous flow chemistry: Automated optimization

towards the Pareto front of multiple objectives," *Chemical Engineering Journal,* vol. 352, pp. 277-282, 2018, doi: 10.1016/j.cej.2018.07.031.

[51] B. Rohr, H. S. Stein, D. Guevarra, Y. Wang, J. A. Haber, M. Aykol, S. K. Suram, and J. M. Gregoire, "Benchmarking the acceleration of materials discovery by sequential learning," *Chemical Science,* vol. 11, no. 10, pp. 2696-2706, 2020, doi: 10.1039/C9SC05999G.

[52] F. Häse, M. Aldeghi, R. J. Hickman, L. M. Roch, M. Christensen, E. Liles, J. E. Hein, and A. Aspuru-Guzik, "Olympus: a benchmarking framework for noisy optimization and experiment planning," *Machine Learning: Science and Technology,* vol. 2, no. 3, 2021, doi: 10.1088/2632-2153/abedc8.

[53] Q. Liang, A. E. Gongora, Z. Ren, A. Tiihonen, Z. Liu, S. Sun, J. R. Deneault, D. Bash, F. Mekki-Berrada, S. A. Khan, K. Hippalgaonkar, B. Maruyama, K. A. Brown, J. F. III, and T. Buonassisi, "Benchmarking the Performance of Bayesian Optimization across Multiple Experimental Materials Science Domains," 2021, arXiv:2106.01309. [Online] Available: arxiv.org/abs/2106.01309

[54] D. E. Graff, E. I. Shakhnovich, and C. W. Coley, "Accelerating high-throughput virtual screening through molecular pool-based active learning," *Chemical Science,* 10.1039/D0SC06805E vol. 12, no. 22, pp. 7866-7881, 2021, doi: 10.1039/D0SC06805E.

[55] W. Ye, C. Chen, Z. Wang, I. H. Chu, and S. P. Ong, "Deep neural networks for accurate predictions of crystal stability," *Nature Communications,* vol. 9, no. 1, p. 3800, 2018, doi: 10.1038/s41467-018-06322-x.

[56] A. Tiihonen, S. J. Cox-Vazquez, Q. Liang, M. Ragab, Z. Ren, N. T. P. Hartono, Z. Liu, S. Sun, C. Zhou, N. C. Incandela, J. Limwongyut, A. S. Moreland, S. Jayavelu, G. C. Bazan, and T. Buonassisi, "Predicting antimicrobial activity of conjugated oligoelectrolyte molecules via machine learning," 2021, arXiv:2105.10236v2. [Online] Available: arxiv.org/abs/2105.10236

[57] Z. Wang, C. Gehring, P. Kohli, and S. Jegelk, "Batched large-scale Bayesian optimization in high-dimensional spaces," 2018, arXiv:1706.01445v4. [Online] Available: arxiv.org/abs/1706.01445

[58] Y.-F. Lim, C. K. Ng, U. S. Vaitesswar, and K. Hippalgaonkar, "Extrapolative Bayesian Optimization with Gaussian Process and Neural Network Ensemble Surrogate Models," *Advanced Intelligent Systems,* no. 2100101, 2021, doi: 10.1002/aisy.202100101.

[59] M. Abdar, F. Pourpanah, S. Hussain, D. Rezazadegan, L. Liu, M. Ghavamzadeh, P. Fieguth, X. Cao, A. Khosravi, U. R. Acharya, V. Makarenkov, and S. Nahavandi, "A Review of Uncertainty Quantification in Deep Learning: Techniques, Applications and Challenges," 2021, arXiv:2011.06225v4 [Online] Available: arxiv.org/abs/2011.06225

[60] T. Pfingsten, "Bayesian active learning for sensitivity analysis," presented at the European Conference on Machine Learning, 2006. [Online]. Available: https://link.springer.com/chapter/10.1007/11871842_35.

[61] L. Hirschfeld, K. Swanson, K. Yang, R. Barzilay, and C. W. Coley, "Uncertainty Quantification Using Neural Networks for Molecular Property Prediction," *Journal of Chemical Information and Modeling,* vol. 60, no. 8, pp. 3770-3780, 2020, doi: 10.1021/acs.jcim.0c00502.

[62] B. Burger, P. M. Maffettone, V. V. Gusev, C. M. Aitchison, Y. Bai, X. Wang, X. Li, B. M. Alston, B. Li, R. Clowes, N. Rankin, B. Harris, R. S. Sprick, and A. I. Cooper, "A mobile robotic chemist," *Nature,* vol. 583, no. 7815, pp. 237-241, 2020, doi:

[63] R. Yuan, Z. Liu, P. V. Balachandran, D. Xue, Y. Zhou, X. Ding, J. Sun, D. Xue, and T. Lookman, "Accelerated discovery of large electrostrains in BaTiO3 -based piezoelectrics using active learning," *Advanced Materials,* vol. 30, no. 7, 2018, doi: 10.1002/adma.201702884.

[64] A. Solomou, G. Zhao, S. Boluki, J. K. Joy, X. Qian, I. Karaman, R. Arróyave, and D. C. Lagoudas, "Multi-objective Bayesian materials discovery: Application on the discovery of precipitation strengthened NiTi shape memory alloys through micromechanical modeling," *Materials & Design,* vol. 160, pp. 810-827, 2018, doi: 10.1016/j.matdes.2018.10.014.

[65] M. Aldeghi, F. Häse, R. J. Hickman, I. Tamblyn, and A. Aspuru-Guzik, "Golem: An algorithm for robust experiment and process optimization," 2021, arXiv:2103.03716. [Online] Available: arxiv.org/abs/2103.03716

[66] M. Demant, P. Virtue, A. Kovvali, S. X. Yu, and S. Rein, "Learning Quality Rating of As-Cut mc-Si Wafers via Convolutional Regression Networks," *IEEE Journal of Photovoltaics,* vol. 9, no. 4, pp. 1064-1072, 2019, doi: 10.1109/JPHOTOV.2019.2906036.

[67] P. Kunze, S. Rein, M. Hemsendorf, K. Ramspeck, and M. Demant, "Learning an empirical digital twin from measurement images for a comprehensive quality inspection of solar cells," *Solar RRL,* vol. Online Version (in press), 2021, doi: 10.1002/solr.202100483.

[68] A. E. Gongora, K. L. Snapp, E. Whiting, P. Riley, K. G. Reyes, E. F. Morgan, and K. A. Brown, "Using simulation to accelerate autonomous experimentation: A case study using mechanics," *iScience,* vol. 24, no. 4, p. 102262, 2021, doi: 10.1016/j.isci.2021.102262.

[69] Z. Li, T. R. Klein, D. H. Kim, M. Yang, J. J. Berry, M. F. A. M. van Hest, and K. Zhu, "Scalable fabrication of perovskite solar cells," *Nature Reviews Materials,* vol. 3, no. 4, p. 18017, 2018, doi: 10.1038/natrevmats.2018.17.

[70] Y. Yang, Z. Xue, L. Chen, C. F. J. Lau, and Z. Wang, "Large-area perovskite films for PV applications: A perspective from nucleation and crystallization," *Journal of Energy Chemistry,* vol. 59, pp. 626-641, 2021, doi: https://doi.org/10.1016/j.jechem.2020.12.001.

[71] M. T. Hoang, F. Ünlü, W. Martens, J. Bell, S. Mathur, and H. Wang, "Towards the environmentally friendly solution processing of metal halide perovskite technology," *Green Chemistry,* 10.1039/D1GC01756J vol. 23, no. 15, pp. 5302-5336, 2021, doi: 10.1039/D1GC01756J.

[72] A. E. Siemenn, E. Shaulsky, M. Beveridge, T. Buonassisia, S. M. Hashmi, and I. Droric, "Autonomous optimization of fluid systems at varying length scales," 2021, arXiv:2015.13553v3. [Online] Available: arxiv.org/abs/2015.13553

[73] "XSolar-Hetero | photovoltaic simulation platform." https://www.xsolar-hetero.sg/, accessed 10/06/2021,

[74] G. Anand, C. Ke, J. Wong, A. Aberle, and R. Stangl, "An online, web based solar cell simulation interface for the personalized simulation of various solar cell architectures, using various simulation programs," in *32th European Photovoltaic Conference on Solar Energy Conversion (EU-PVSEC)* 2016, doi: 10.13140/RG.2.1.3289.7521.

[75] Y. Zhao, K. Zhan, Z. Wang, and W. Shen, "Deep learning-based automatic detection of multitype defects in photovoltaic modules and application in real production line," *Progress in Photovoltaics: Research and Applications,* vol. 29, no. 4, pp. 471-484, 2021,

doi: 10.1002/pip.3395.
[76] L. Bommes, T. Pickel, C. Buerhop-Lutz, J. Hauch, C. Brabec, and I. M. Peters, "Computer vision tool for detection, mapping, and fault classification of photovoltaics modules in aerial IR videos," *Progress in Photovoltaics: Research and Applications,* vol. n/a, no. n/a, 2021, doi: 10.1002/pip.3448.
[77] Y. Buratti, Z. Abdullah-Vetter, A. Sowmya, T. Trupke, and Z. Hameiri, "A Deep Learning Approach for Loss-Analysis from Luminescence Images," in *48th IEEE Photovoltaic Specialists Conference (PVSC)*, 20-25 June 2021 2021, pp. 0097-0100, doi: 10.1109/PVSC43889.2021.9518512. [Online]. Available: https://ieeexplore.ieee.org/document/9518512/
[78] S. I. P. Tian, Z. Liu, V. Chellappan, Y. F. Lim, Z. Ren, F. Oviedo, B. H. Teo, J. Thapa, R. Dutta, B. P. MacLeod, F. G. L. Parlane, J. Senthilnath, C. P. Berlinguette, and T. Buonassisi, "Rapid and Accurate Thin Film Thickness Extraction via UV-Vis and Machine Learning," presented at the 47th IEEE Photovoltaic Specialists Conference (PVSC), 15 June-21 Aug, 2020. [Online]. Available: https://ieeexplore.ieee.org/document/9300634/.
[79] N. Taherimakhsousi, B. P. MacLeod, F. G. L. Parlane, T. D. Morrissey, E. P. Booker, K. E. Dettelbach, and C. P. Berlinguette, "Quantifying defects in thin films using machine vision," *npj Computational Materials,* vol. 6, no. 1, 2020, doi: 10.1038/s41524-020-00380-w.
[80] R. E. Brandt, R. C. Kurchin, V. Steinmann, D. Kitchaev, C. Roat, S. Levcenco, G. Ceder, T. Unold, and T. Buonassisi, "Rapid photovoltaic device characterization through bayesian parameter estimation," *Joule,* vol. 1, no. 4, pp. 843-856, 2017, doi: 10.1016/j.joule.2017.10.001.
[81] R. C. Kurchin, J. R. Poindexter, V. Vähänissi, H. Savin, C. d. Cañizo, and T. Buonassisi, "How Much Physics is in a Current–Voltage Curve? Inferring Defect Properties From Photovoltaic Device Measurements," *IEEE Journal of Photovoltaics,* vol. 10, no. 6, pp. 1532-1537, 2020, doi: 10.1109/JPHOTOV.2020.3010105.
[82] Z. Ren, F. Oviedo, M. Thway, S. I. P. Tian, Y. Wang, H. Xue, J. Dario Perea, M. Layurova, T. Heumueller, E. Birgersson, A. G. Aberle, C. J. Brabec, R. Stangl, Q. Li, S. Sun, F. Lin, I. M. Peters, and T. Buonassisi, "Embedding physics domain knowledge into a Bayesian network enables layer-by-layer process innovation for photovoltaics," *npj Computational Materials,* vol. 6, no. 1, 2020, doi: 10.1038/s41524-020-0277-x.
[83] F. Oviedo, "Accelerated development of photovoltaics by physics-informed machine learning," Ph.D., Department of Mechanical Engineering, Massachusetts Institute of Technology, https://dspace.mit.edu/handle/1721.1/127060, 2020.
[84] V. M. Le Corre, M. Stolterfoht, L. Perdigón Toro, M. Feuerstein, C. Wolff, L. Gil-Escrig, H. J. Bolink, D. Neher, and L. J. A. Koster, "Charge Transport Layers Limiting the Efficiency of Perovskite Solar Cells: How To Optimize Conductivity, Doping, and Thickness," *ACS Applied Energy Materials,* vol. 2, no. 9, pp. 6280-6287, 2019, doi: 10.1021/acsaem.9b00856.
[85] V. M. Le Corre, Z. Wang, L. J. A. Koster, and W. Tress, "Device Modeling of Perovskite Solar Cells: Insights and Outlooks," in *Soft-Matter Thin Film Solar Cells*, 2020, ch. 4, pp. 1-32.
[86] P. Lopez-Varo, J. A. Jiménez-Tejada, M. García-Rosell, S. Ravishankar, G. Garcia-Belmonte, J. Bisquert, and O. Almora, "Device Physics of Hybrid Perovskite Solar cells:

[87] N. Tessler and Y. Vaynzof, "Insights from Device Modeling of Perovskite Solar Cells," *ACS Energy Letters,* vol. 5, no. 4, pp. 1260-1270, 2020, doi: 10.1021/acsenergylett.0c00172.

[88] N. T. P. Hartono, J. Thapa, A. Tiihonen, F. Oviedo, C. Batali, J. J. Yoo, Z. Liu, R. Li, D. F. Marrón, M. G. Bawendi, T. Buonassisi, and S. Sun, "How machine learning can help select capping layers to suppress perovskite degradation," *Nature Communications,* vol. 11, no. 1, p. 4172, 2020, doi: 10.1038/s41467-020-17945-4.

[89] S. G. Hashmi, A. Tiihonen, D. Martineau, M. Ozkan, P. Vivo, K. Kaunisto, V. Ulla, S. M. Zakeeruddin, and M. Grätzel, "Long term stability of air processed inkjet infiltrated carbon-based printed perovskite solar cells under intense ultra-violet light soaking," *Journal of Materials Chemistry A,* vol. 5, no. 10, pp. 4797-4802, 2017, doi: 10.1039/C6TA10605F.

[90] R. J. Stoddard, W. A. Dunlap-Shohl, H. Qiao, Y. Meng, W. F. Kau, and H. W. Hillhouse, "Forecasting the Decay of Hybrid Perovskite Performance Using Optical Transmittance or Reflected Dark-Field Imaging," *ACS Energy Letters,* vol. 5, no. 3, pp. 946-954, 2020, doi: 10.1021/acsenergylett.0c00164.

[91] J. M. Howard, Q. Wang, E. Lee, R. Lahoti, T. Gong, M. Srivastava, A. Abate, and M. S. Leite, "Quantitative predictions of photo-emission dynamics in metal halide perovskites via machine learning," 2020, arXiv:2010.03702. [Online] Available: arxiv.org/abs/2010.03702

[92] C. Gottbrath, J. Bailin, C. Meakin, T. Thompson, and J. J. Charfman, "The Effects of Moore's Law and Slacking on Large Computations," 1999, arXiv:astro-ph/9912202v1 [Online] Available: arxiv.org/abs/astro-ph/9912202

[93] "Machine Learning Engineering for Production (MLOps) Specialization." https://www.coursera.org/specializations/machine-learning-engineering-for-production-mlops, accessed 10/06/2021,

[94] www.youtube.com/channel/UCxaokYYzFI9XPOUP_W_sD9g, accessed 10/03/2021,

[95] "LinkedIn Group: Autonomous Research Systems." www.linkedin.com/groups/12176428/, accessed 10/06/2021,

[96] J. M. Utterback, *Mastering the Dynamics of Innovation*. Harvard Business Review Press, 1994.

[97] "GitHub - GPyOpt." github.com/SheffieldML/GPyOpt/blob/master/README.md accessed 10/06/2021,

[98] M. Balandat, B. Karrer, D. R. Jiang, S. Daulton, B. Letham, A. G. Wilson, and E. Bakshy, "BoTorch: A Framework for Efficient Monte-Carlo Bayesian Optimization," 2019, arXiv:1910.06403v3 [Online] Available: arxiv.org/abs/1910.06403

[99] E. I. Marchenko, S. A. Fateev, A. A. Petrov, V. V. Korolev, A. Mitrofanov, A. V. Petrov, E. A. Goodilin, and A. B. Tarasov, "Database of two-dimensional hybrid perovskite materials: open-access collection of crystal structures, band gaps, and atomic partial charges predicted by machine learning," *Chemistry of Materials,* vol. 32, no. 17, pp. 7383-7388, 2020, doi: 10.1021/acs.chemmater.0c02290.

[100] Y. Cai, W. Xie, Y. T. Teng, P. C. Harikesh, B. Ghosh, P. Huck, K. A. Persson, N. Mathews, S. G. Mhaisalkar, M. Sherburne, and M. Asta, "High-throughput computational study of halide double perovskite inorganic compounds," *Chemistry of Materials,* vol. 31,

no. 15, pp. 5392-5401, 2019, doi: 10.1021/acs.chemmater.9b00116.

[101] C. Kim, G. Pilania, and R. Ramprasad, "Machine Learning Assisted Predictions of Intrinsic Dielectric Breakdown Strength of ABX3 Perovskites," *The Journal of Physical Chemistry C,* vol. 120, no. 27, pp. 14575-14580, 2016, doi: 10.1021/acs.jpcc.6b05068.

[102] A. Zunger, "Inverse design in search of materials with target functionalities," *Nature Reviews Chemistry,* vol. 2, no. 4, p. 0121, 2018, doi: 10.1038/s41570-018-0121.

[103] A. Franceschetti and A. Zunger, "The inverse band-structure problem of finding an atomic configuration with given electronic properties," *Nature,* vol. 402, no. 6757, pp. 60-63, 1999, doi: 10.1038/46995.

[104] R. Gómez-Bombarelli, J. N. Wei, D. Duvenaud, J. M. Hernández-Lobato, B. Sánchez-Lengeling, D. Sheberla, J. Aguilera-Iparraguirre, T. D. Hirzel, R. P. Adams, and A. Aspuru-Guzik, "Automatic Chemical Design Using a Data-Driven Continuous Representation of Molecules," *ACS Central Science,* vol. 4, no. 2, pp. 268-276, 2018, doi: 10.1021/acscentsci.7b00572.

[105] Z. Ren, J. Noh, S. Tian, F. Oviedo, G. Xing, Q. Liang, A. Aberle, Y. Liu, Q. Li, S. Jayavelu, K. Hippalgaonkar, Y. Jung, and T. Buonassisi, "Inverse design of crystals using generalized invertible crystallographic representation," 2020, arXiv:2005.07609v2 [Online] Available: arxiv.org/abs/2005.07609

[106] R.-R. Griffiths and J. M. Hernández-Lobato, "Constrained Bayesian optimization for automatic chemical design using variational autoencoders," *Chemical Science,* vol. 11, no. 2, pp. 577-586, 2020, doi: 10.1039/C9SC04026A.

[107] K. Sattari, Y. Xie, and J. Lin, "Data-driven algorithms for inverse design of polymers," *Soft Matter,* vol. 17, no. 33, pp. 7607-7622, 2021, doi: 10.1039/d1sm00725d.

[108] M. J. McDermott, S. S. Dwaraknath, and K. A. Persson, "A graph-based network for predicting chemical reaction pathways in solid-state materials synthesis," *Nature Communications,* vol. 12, no. 1, p. 3097, 2021, doi: 10.1038/s41467-021-23339-x.

[109] W. Ye, C. Chen, Z. Wang, I.-H. Chu, and S. P. Ong, "Deep neural networks for accurate predictions of crystal stability," *Nature Communications,* vol. 9, no. 1, p. 3800, 2018, doi: 10.1038/s41467-018-06322-x.

[110] A. Vasylenko, J. Gamon, B. B. Duff, V. V. Gusev, L. M. Daniels, M. Zanella, J. F. Shin, P. M. Sharp, A. Morscher, R. Chen, A. R. Neale, L. J. Hardwick, J. B. Claridge, F. Blanc, M. W. Gaultois, M. S. Dyer, and M. J. Rosseinsky, "Element selection for crystalline inorganic solid discovery guided by unsupervised machine learning of experimentally explored chemistry," *Nature Communications,* vol. 12, no. 1, p. 5561, 2021, doi: 10.1038/s41467-021-25343-7.

[111] W. Sun, S. T. Dacek, S. P. Ong, G. Hautier, A. Jain, W. D. Richards, A. C. Gamst, K. A. Persson, and G. Ceder, "The thermodynamic scale of inorganic crystalline metastability," *Science Advances,* vol. 2, no. 11, p. e1600225, 2016, doi: doi:10.1126/sciadv.1600225.

[112] J. Jang, G. H. Gu, J. Noh, J. Kim, and Y. Jung, "Structure-Based Synthesizability Prediction of Crystals Using Partially Supervised Learning," *Journal of the American Chemical Society,* vol. 142, no. 44, pp. 18836-18843, 2020, doi: 10.1021/jacs.0c07384.

[113] E. Kim, K. Huang, A. Saunders, A. McCallum, G. Ceder, and E. Olivetti, "Materials Synthesis Insights from Scientific Literature via Text Extraction and Machine Learning," *Chemistry of Materials,* vol. 29, no. 21, pp. 9436-9444, 2017, doi: 10.1021/acs.chemmater.7b03500.

[114] R. Roscher, B. Bohn, M. F. Duarte, and J. Garcke, "Explainable Machine Learning for

Scientific Insights and Discoveries," *IEEE Access,* vol. 8, pp. 42200-42216, 2020, doi: 10.1109/ACCESS.2020.2976199.

[115] A. Barredo Arrieta, N. Díaz-Rodríguez, J. Del Ser, A. Bennetot, S. Tabik, A. Barbado, S. Garcia, S. Gil-Lopez, D. Molina, R. Benjamins, R. Chatila, and F. Herrera, "Explainable Artificial Intelligence (XAI): Concepts, taxonomies, opportunities and challenges toward responsible AI," *Information Fusion,* vol. 58, pp. 82-115, 2020, doi: 10.1016/j.inffus.2019.12.012.

[116] R. R. Hoffman, S. T. Mueller, G. Klein, and J. Litman, "Metrics for Explainable AI: Challenges and Prospects," 2019, arXiv:1812.04608v2. [Online] Available: arxiv.org/abs/1812.04608

[117] H. Wagner-Mohnsen and P. P. Altermatt, "A Combined Numerical Modeling and Machine Learning Approach for Optimization of Mass-Produced Industrial Solar Cells," *IEEE Journal of Photovoltaics,* vol. 10, no. 5, pp. 1441-1447, 2020, doi: 10.1109/JPHOTOV.2020.3004930.

[118] J. Pan, K. L. Low, J. Ghosh, S. Jayavelu, M. M. Ferdaus, S. Y. Lim, E. Zamburg, Y. Li, B. Tang, X. Wang, J. F. Leong, S. Ramasamy, T. Buonassisi, C.-K. Tham, and A. V.-Y. Thean, "Transfer Learning-Based Artificial Intelligence-Integrated Physical Modeling to Enable Failure Analysis for 3 Nanometer and Smaller Silicon-Based CMOS Transistors," *ACS Applied Nano Materials,* vol. 4, no. 7, pp. 6903-6915, 2021, doi: 10.1021/acsanm.1c00960.

[119] S. Liu, B. B. Kappes, B. Amin-ahmadi, O. Benafan, X. Zhang, and A. P. Stebner, "Physics-informed machine learning for composition – process – property design: Shape memory alloy demonstration," *Applied Materials Today,* vol. 22, p. 100898, 2021, doi: 10.1016/j.apmt.2020.100898.

[120] L. S. Shapley, "Contributions to the Theory of Games," no. 2,, p. 307, 1953.

[121] S. M. Lundberg and S.-I. Lee, "A unified approach to interpreting model predictions," in *Proceedings of the 31st international conference on neural information processing systems (NeurIPS)*, 2017, pp. 4768-4777.

[122] M. T. Ribeiro, S. Singh, and C. Guestrin, ""Why Should I Trust You?"," in *the 22nd ACM SIGKDD International Conference on Knowledge Discovery and Data Mining*, New York, NY, USA, 2016 2016: ACM, doi: 10.1145/2939672.2939778.

[123] I. E. Kumar, S. Venkatasubramanian, C. Scheidegger, and S. Friedler, "Problems with Shapley-value-based explanations as feature importance measures," 2020, arXiv:2002.11097v2. [Online] Available: arxiv.org/abs/2002.11097

[124] R. L. Phillips, K. H. Chang, and S. A. Friedler, "Interpretable Active Learning," 2018, arXiv:1708.00049v2 [Online] Available: arxiv.org/abs/1708.00049

[125] D. Slack, S. A. Friedler, C. Scheidegger, and C. D. Roy, "Assessing the Local Interpretability of Machine Learning Models," 2019, arxiv: 1902.03501v2. [Online] Available: arxiv.org/abs/1902.03501

[126] C. S. Kong, W. Luo, S. Arapan, P. Villars, S. Iwata, R. Ahuja, and K. Rajan, "Information-theoretic approach for the discovery of design rules for crystal chemistry," *Journal of Chemical Information and Modeling,* vol. 52, no. 7, pp. 1812-1820, 2012, doi: 10.1021/ci200628z.

[127] R. R. Naik, A. Tiihonen, J. Thapa, C. Batali, Z. Liu, S. Sun, and T. Buonassisi, "Discovering Equations that Govern Experimental Materials Stability under Environmental Stress using Scientific Machine Learning," 2021, arXiv:2106.10951.

[Online] Available: arxiv.org/abs/2106.10951

[128] S. H. M. Mehr, M. Craven, A. I. Leonov, G. Keenan, and L. Cronin, "A universal system for digitization and automatic execution of the chemical synthesis literature," *Science,* vol. 370, no. 6512, pp. 101-108, 2020, doi: 10.1126/science.abc2986.

[129] M. C. Swain and J. M. Cole, "ChemDataExtractor: A Toolkit for Automated Extraction of Chemical Information from the Scientific Literature," *Journal of Chemical Information and Modeling,* vol. 56, no. 10, pp. 1894-1904, 2016, doi: 10.1021/acs.jcim.6b00207.

[130] L. Hawizy, D. M. Jessop, N. Adams, and P. Murray-Rust, "ChemicalTagger: A tool for semantic text-mining in chemistry," *Journal of Cheminformatics,* vol. 3, no. 1, pp. 17-17, 2011, doi: 10.1186/1758-2946-3-17.

[131] E. Schwenker, W. Jiang, T. Spreadbury, N. Ferrier, O. Cossairt, and M. K. Y. Chan, "EXSCLAIM!--An automated pipeline for the construction of labeled materials imaging datasets from literature," 2021, arXiv:2103.10631. [Online] Available: arxiv.org/abs/2103.10631